\documentclass[twocolumn,showpacs]{revtex4}
\usepackage{graphicx}
\usepackage{dcolumn}
\usepackage{amsmath}

\begin{document}

\preprint{sway3}

\title[Short Title]{Posture Noise and the Measured Probability to Fall}

\author{Radostina K. Koleva}
\affiliation{Electrical and Computer Engineering Department, 
University of Maryland, College Park MD 20742} 
\author{A. Widom} 
\author{D. Garelick}
\affiliation{Physics Department, Northeastern University, 
Boston MA 02115}
\author{Meridith Harris}
\author{Bryan A. Spinelli}
\author{Amanda E. Finn}
\author{Kellie Bonner}
\author{S. White}
\author{Vijay Daryanana}
\author{P. Nyatsambo}
\author{M. Patel}
\author{R. Sampson}
\affiliation{Department of Physical Therapy, 
Northeastern University, Boston MA 02115}

\date{\today}

\begin{abstract}
Experimental evidence is presented connecting small fluctuations 
in the posture of a quiet standing subject and the probability that 
the subject will have an accidental fall within a time period 
of one year. The data can be understood on the basis of random 
velocity fluctuations providing kinetic energy in dynamical 
posture modes. The kinetic energy can activate a transition 
over the a potential energy barrier which keeps subjects 
in a metastable standing mode. The probability for a fall 
then follows an Arrhenius activation law.
\end{abstract}


\maketitle

\section{Introduction}

In recent years there has been considerable interest in the nature of posture 
sway~\cite{alonso00, chow95, collins94, collins95, thurner00, lauk98, yao01}.
Sway is the name given to the small 
dynamic displacement fluctuations present in the posture coordinates 
of a person who is otherwise standing quietly. Some studies have been 
motivated (in part) by a practical medical problem, i.e. 
the consequences of falling for the elderly population~\cite{collins95B, 
firsov93, rosenblum98}. Such falls are frequent events. Medical studies 
show that  \begin{math}\sim 40\% \end{math} of the elderly 
population experience a fall at least once a year causing fractures, 
contusions, head injuries, joint distortions and 
dislocations~\cite{gregg00, kannus00}. The medical costs in the United 
Sates are  \begin{math}\sim 10^7 \$/{\rm year} \end{math}
~\cite{gregg00, sherrington98}.  
These figures suggest the desirability of a reliable screening 
technique for determining balance impairment, especially among the 
elderly. Our purpose is to present experimental evidence concerning 
the connection between noise processes in posture sway and the 
probability of falling. Some theoretical notions concerning this 
connection have been previously discussed~\cite{koleva01}.
 
In Sec.II, the experimental methods employed in measuring posture 
sway coordinates are discussed. The central instrument involved 
in the measurement is a sound wave assessment device (SWA). 
In Sec.III, the free energy barrier model~\cite{koleva01} 
for computing the probability 
of a fall is discussed. The most important posture parameter determining 
the probability of a fall is the root-mean-square (rms) velocity 
\begin{math} v \end{math} of the fluctuating anteroposterior 
(back and forth) posture oscillation mode. The probability for a fall 
is to be described by the function 
\begin{equation}
p(v)=e^{-(u/v)^2},
\end{equation}
where \begin{math} u^2  \end{math} is proportional to the height of 
an energy barrier preventing a fall. Eq.(1) will be simply derived in 
what follows, but here one may note that \begin{math} u \end{math} 
gives rise to a single parameter fit to the data on posture sway 
and the probability of a fall. A maximum likelihood analysis of 
\begin{math} u  \end{math} from experimental data is discussed in 
Sec.IV. In the concluding Sec.V, we discuss the utility of the 
energy barrier view in predicting the fallers within the elderly 
population. 

\section{Experimental Procedure}

The SWA device consists of two small ultrasonic transducers of the type 
used by Polaroid for auto-focusing cameras. Each transducer is able to 
emit and detect ultrasonic pulses at \begin{math} 30\ {\rm Hz}\end{math}. 
When positioned some distance apart, say  
\begin{math}L\approx 170\ {\rm cm}\end{math}, 
the transducers are able to determine the distance between 
them by measuring the time it takes for a pulse to propagate 
from one transducer to the other. The position accuracy of such a 
measurement is \begin{math}\Delta x \approx 0.02 cm \end{math}. 

The above technique is well suited for the measurement of the fine 
movements of subjects during quiet standing. One transducer is 
attached to the lower back (around the waist line) of a subject. 
The other transducer is positioned on a stable laboratory stand. 
The duration of the quiet standing measurement is 60 seconds.

The subjects participating in this study were 56 mature adults of ages
from \begin{math} 60\ {\rm years}\end{math} to 
\begin{math} 97\ {\rm years} \end{math} 
residing in independent community residential facilities. 
The subjects filled out a questionnaire in order to determine their 
medical history and their history of falls. The subjects included 
in the study all had good ability to follow directions, 
good visual activity, good hearing and the ability to stand 
without assistance for at least 60 seconds.  Excluded
were subjects who had unstable cardiac diseases, orthostatic
hypertension, severe neurological diseases, blindness, diabetes, 
mental diseases, or lower extremities amputations. 

The subjects were asked to stand for one minute while their movement
(back and forth) in the anteroposterior direction was measured - i.e. 
the posture coordinate displacement \begin{math} x(t) \end{math} 
was recorded. The time interval \begin{math} \Delta t \end{math}  
between displacement measurements was  
\begin{math} \Delta t^{-1}=30\ {\rm Hz} \end{math}. 
The rms velocity \begin{math} v=\sqrt{<\dot{x}^2>} \end{math} 
was determined from the measured displacements and time 
intervals between displacement measurements.

The measured rms velocity \begin{math} v \end{math} was recorded 
within a bin of size  
\begin{math} \Delta v=0.05\ {\rm cm/sec} \end{math}.
In the \begin{math} k^{th} \end{math} data bin of rms velocity 
\begin{math} |v-v_k|<(\Delta v/2) \end{math}, 
\begin{math} N_k \end{math} subjects were studied. Of these,  
\begin{math} n_k\le N_k \end{math} subjects indicated that they 
had a fall within the last year. The measured probability for 
falling within a year was taken to be the fraction 
\begin{math} p_k=(n_k/N_k) \end{math}.

\section{Theoretical Model}

\begin{figure}[bp]
\scalebox {0.5}{\includegraphics{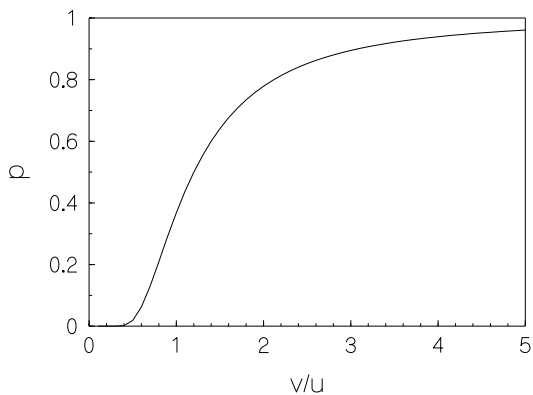}}
\caption{Shown is the theoretical probability {\em p} for a fall 
as a function of the rms velocity {\em v} as given in Eq.(1).}
\label{fig1}
\end{figure}

The physical kinetics of thermally activated processes is often 
modeled using a free energy of activation 
\begin{math} \phi  \end{math} and a temperature 
\begin{math} T \end{math}. The Arrhenius law for the probability 
\begin{math} p \end{math}
of overcoming a local activation barrier and falling from a meta-stable 
free energy to a lower stable free energy is given by 
\begin{equation}
p=e^{-\phi /k_BT}.
\end{equation}
The temperature in the Arrhenius law represents the thermal 
rms velocity \begin{math} v  \end{math} of the coordinate 
describing the free energy barrier; i.e. for a coordinate associated 
with a mass \begin{math} \mu \end{math}, the equipartition theorem of 
statistical mechanics asserts that 
\begin{equation}
\mu v^2=k_BT.
\end{equation} 
Expressing the energy barrier in terms of a ``velocity''  
\begin{math} u \end{math} via 
\begin{equation}
\mu u^2=\phi 
\end{equation}
yields 
\begin{equation}
p=e^{-u^2/v^2}.
\end{equation}

Our model Eq.(1) is merely Eqs.(2) and (5) in thinly disguised form. 
However, the following physical points are worthy of note: 
(i) The ``noise temperature'' \begin{math} T_n  \end{math} (with 
\begin{math} \mu v^2=k_BT_n  \end{math}) of living beings 
is {\em not} the environmental temperature \begin{math} T \end{math}. 
An object in thermal equilibrium with its environment 
is dead\cite{kubo99}. The measured rms velocity 
\begin{math} v \end{math} of the living subjects represents 
noise but not thermal noise. (ii) The velocity 
\begin{math} u \end{math} (assumed independent of  noise  
temperature) is still determined by the energy barrier 
which prevents us from falling without (say) being pushed. 

The potential energy and energy barriers, at various measured 
posture coordinate \begin{math} x \end{math} values, describes 
a combination of motions of the ankles and 
hips~\cite{nashner85, shumway95, whittle91, winter90}. The motions 
(referred to as ``strategies'' in the biomedical literature) are 
familiar to  many casual observers. A loss of balance may often be 
preceded by large amplitude rocking back and forth  
(without stepping protections) before the final fall actually 
takes place.

The theoretical probability for a fall as a function of rms 
velocity is shown in FIG. 1. Note that the probability for a fall 
becomes appreciable when the rms velocity \begin{math} v \end{math}
is somewhat greater than half the value of \begin{math} u \end{math}. 
The experimental determination of \begin{math} u \end{math} is thereby 
of great interest.  

\section{Data Analysis}

\begin{figure}[bp]
\scalebox {0.5}{\includegraphics{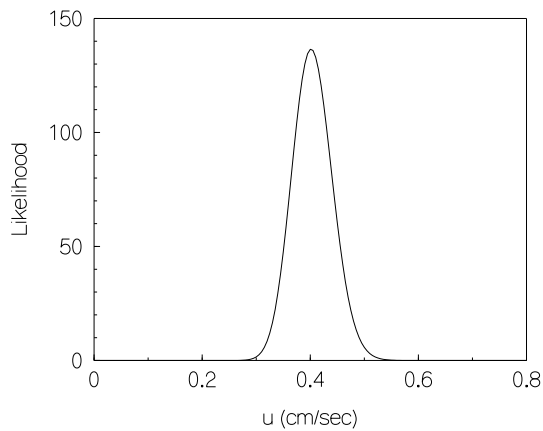}}
\caption{Shown is the experimental likelihood {\em L(u)} 
of finding a barrier to a fall described by the parameter 
{\em u}.}
\label{fig2}
\end{figure}

\begin{figure}[bp]
\scalebox {0.5}{\includegraphics{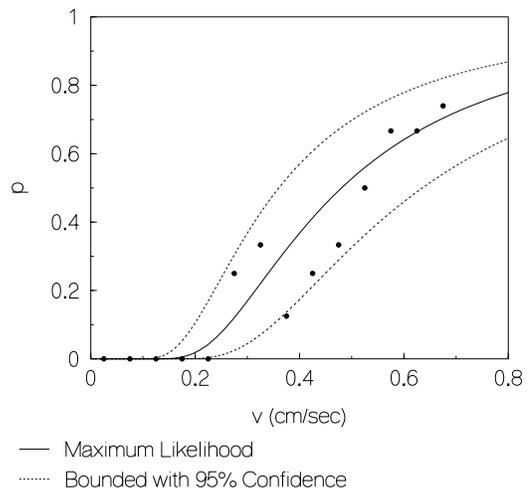}}
\caption{Shown as a solid curve is the best theoretical fit 
to the probability data points found from the maximum likelihood.  
The theoretical probability for a fall is in reasonable  
agreement with the experimental results. The dashed curves 
represent both the upper and lower bounds to the probability 
for a fall at a {\em 95\%} confidence level.}
\label{fig3}
\end{figure}

For \begin{math} \{N_k\} \end{math} subjects having an rms velocities  
\begin{math} \{v_k\}  \end{math}, the total probability of having 
\begin{math} \{n_k<N_k\}  \end{math} falling subjects is theoretically 
given by the binomial distribution 
\begin{equation}
W=\prod_k \left\{C(N_k,n_k)
p(v_k)^{n_k}
\left(1-p(v_k)\right)^{(N_k-n_k)}\right\},
\end{equation}
where 
\begin{equation}
C(N_k,n_k)=\left({N_k!\over n_k!(N_k-n_k)!}\right).
\end{equation}

The likelihood of a particular value of the parameter 
\begin{math} u \end{math} is obtained from the probability 
distribution \begin{math} W \end{math} in Eq.(6) by inserting 
the experimental data 
\begin{math} \{n_k^{(data)},N_k^{(data)}\} \end{math} 
into 
\begin{equation}
L(u)=K W\left[\{n_k^{(data)},N_k^{(data)}\};u\right], 
\end{equation}
where \begin{math} K  \end{math} is an arbitrary constant. 

The likelihood function \begin{math} L(u) \end{math} for the data at 
hand is plotted in FIG. 2. The likelihood 
\begin{math} L(u) \end{math} has a well defined peak at 
\begin{math} u_{\rm (best\ fit)}=0.40\ {\rm cm/sec}\end{math} which 
(in this analysis) is the best value of \begin{math} u \end{math}   
for fitting the experimental data. 
The likelihood peak gives rise to a clear picture of the uncertainty 
in the value of \begin{math} u \end{math}. If one draws a horizontal 
line on the likelihood plot with \begin{math} c\% \end{math} of the area 
under the likelihood peak being above the horizontal line, 
then one obtains upper and lower bounds 
to \begin{math} u \end{math} by the two intersections of the line 
with the likelihood peak. The upper and lower bounds are thereby 
estimated with a confidence level of \begin{math} c\% \end{math}.

Shown in FIG. 3 are three plots of the probability for a fall, 
together with the experimental points  
\begin{equation}
p_k^{experiment}=\left(n_k^{(data)}/N_k^{(data)}\right).
\end{equation}
The solid curve is the fit of the theoretical ``probability of 
a fall'' curve in FIG. 1 with the experimental data points employing  
the value of \begin{math} u  \end{math} determined by the 
maximum likelihood in FIG 2. The upper and lower bounds are shown 
by dashed curves also obtained from the likelihood curve at a 
confidence level of \begin{math} 95\%  \end{math}. Given the number of 
tested subjects, the agreement between theory and experiment is 
satisfactory. Higher numbers of subjects would be expected to improve 
the analysis by somewhat narrowing the likelihood peak. 

\section{Conclusion}

Experimental evidence has been presented relating the rms 
velocity \begin{math} v \end{math} of the anteroposterior 
posture oscillation mode to the probability of a fall. 
In terms of the noise temperature of this posture mode 
\begin{equation}
k_BT_n=\mu v^2,
\end{equation}
the data can be reasonably understood on the basis of random 
velocity fluctuations providing kinetic energy. The kinetic energy 
can then activate a transition over that potential energy barrier 
which normally keeps subjects in a metastable standing mode.
The probability for a fall then follows an activation law 

While the single parameter \begin{math} u \end{math} gives 
a reasonable  picture of the probability of a fall as shown in FIG. 3, 
several other parameters have been examined but play a minor role 
in determining the probability for a fall. These other factors 
include the expressed fear of falling, self perception of balancing 
ability, the mass body index and muscle strength. 

The existence of an energy barrier to a fall is susceptible to further 
experiments. If one leans forward or backward beyond a 
certain critical displacement, then balance is lost unless one 
employs the step strategy. By measuring the potential energy 
change required to reach this critical displacement, the quantity 
\begin{math} u \end{math} can be estimated. 

Finally, this project was started in order to design an instrument 
which could identify fallers within an elderly community. 
It appears that the SWA device is useful in achieving this goal.

\begin{acknowledgments}
The authors would like to thank Carlos Tun for his contributions 
to the early stages of this project.
\end{acknowledgments}

\end{document}